\documentclass[twocolumn,amsmath,amssymb]{snp}
\usepackage{graphicx}
\usepackage{dcolumn}
\usepackage{bm}
\begin{document}

\title{{\Large Comprehending Particle Production in Proton+Proton and Heavy-ion collisions at the Large Hadron Collider}}

\author{\large Raghunath Sahoo}
\email{Raghunath.Sahoo@cern.ch}
\affiliation{Discipline of Physics, School of Basic Sciences, Indian Institute of Technology Indore, Simrol, Indore - 453552, INDIA}
\maketitle

\section*{Introduction}
In the extreme conditions of temperature and energy density, nuclear matter undergoes a transition to a new phase, which is governed
by partonic degrees of freedom. This phase is called Quark-Gluon Plasma (QGP). The transition to QGP phase was conjectured to take place 
in central nucleus-nucleus collisions \cite{Heiselberg:2000fk}. With the advent of unprecedented collision energy at the Large Hadron Collider (LHC), at CERN, it has been possible to create energy densities higher than that was predicted by lattice QCD for a deconfinement transition, {\it i.e.} 1 $\rm GeV/fm^3$, both in peripheral heavy-ion (A+A) and proton+proton ($p+p$) collisions. $p+p$ collisions show a comparable degree of strangeness enhancement \cite{ALICE-NP} and collectivity, which are the signatures of QGP. Thus it becomes prudent to ask questions like: i) Can we observe deconfinement transition in $p+p$ collisions at the LHC energies?, ii) What is/are the particle production mechanism(s) in $p+p$ and A+A collisions?, iii) Is it possible to relate the observables in $p+p$ and A+A collisions? To address these questions, we proceed as follows to look into some of the important observables related to particle production mechanism, and review the experimental and theoretical results thereof.

\begin{figure}
\includegraphics[width=45mm]{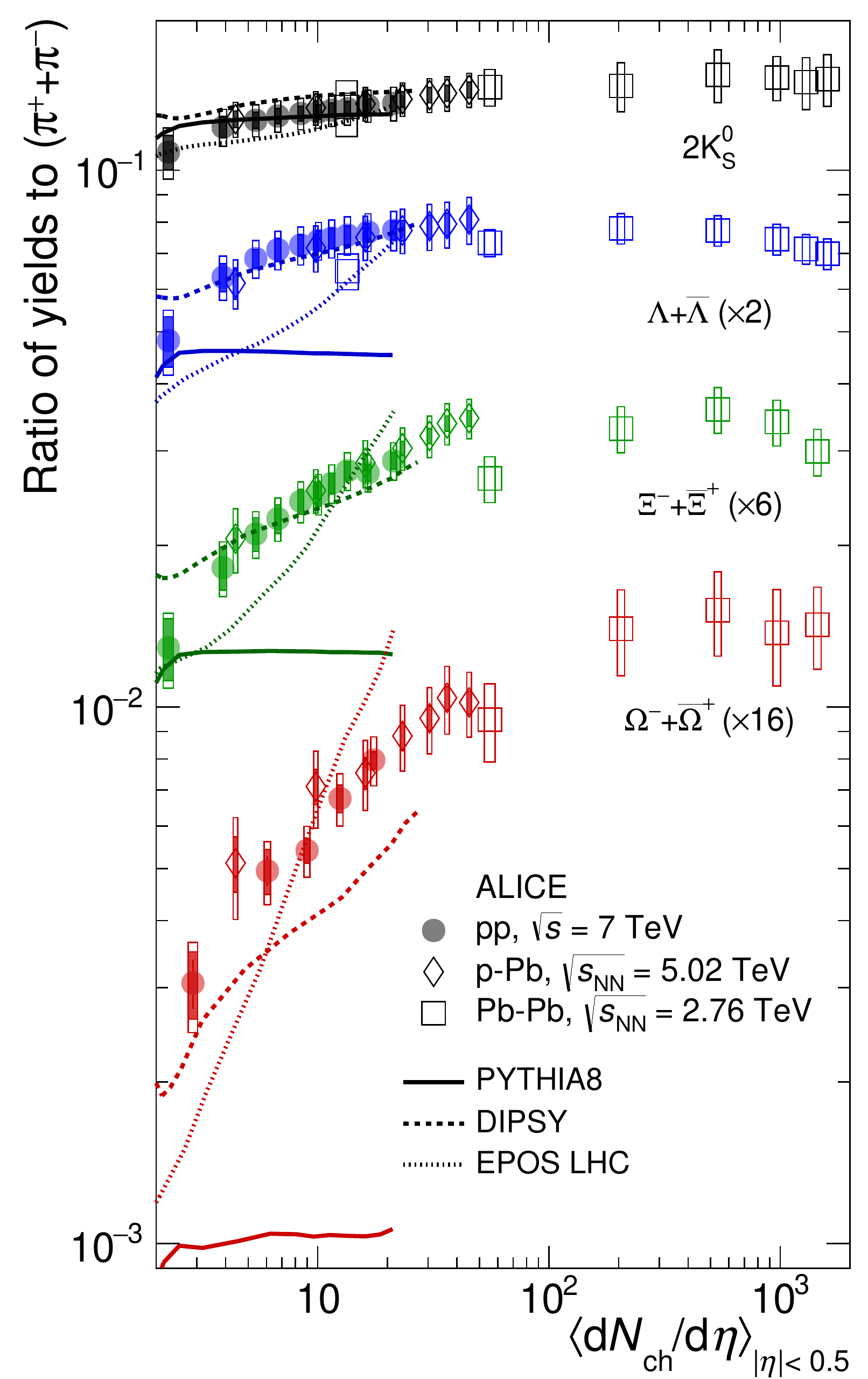}
\caption{\label{fig1}  The yield ratios of strange and multi-strange particle to $(\pi^++\pi^-)$ as a function of $(dN_{\rm ch}/d\eta)$ in $|y| <$ 0.5 \cite{ALICE-NP}.}
\end{figure}

\section*{Transverse Momentum Spectra}
Transverse momentum ($p_T$)-spectra in A+A collisions follow an equilibrium Boltzmann-Gibbs statistics, giving information about, particle 
yield, kinetic freeze-out temperature and radial flow. On the other hand, $p_T$-spectra in $p+p$ collisions are well described by a non-extensive Tsallis distribution, which follows a low-$p_T$ Boltzmann type of distribution and high-$p_T$ power-law behaviour.  This empirical behaviour was earlier proposed by Hagedorn. We have extensively studied the $p_T$-spectra of light flavors and multi-strange particles in $p+p$ collisions at LHC energies. The power-law part in the spectra is responsible for the correlation of the final state particles. Thus the high-multiplicity $p+p$ events seem to be more interesting. The increase in particle multiplicity drives the system towards a thermodynamic equilibrium. A mass dependent differential freeze-out scenario is also observed, meaning high mass particles appear to freeze-out early in time, as compared to lighter ones. $<p_T>$ as a function of charged particle multiplicity ($N_{\rm ch}$) shows a saturation behaviour towards higher
$N_{\rm ch}$-values for A+A collisions, whereas for $p+p$ collisions, it grows up monotonically. This behaviour has been shown to be because of
the multi-partonic interactions and color reconnection mechanisms, which becomes prevalent in high-multiplicity events. 
\section*{Multi-Partonic Interactions (MPI)}
The final state charged particle production in A+A collisions seems to deviate from a nucleon-participant scaling. Several scaling behaviours, like number of
binary nucleon-nucleon collisions, the quark-participant scaling are explored to explain the particle production. In high-multiplicity $p+p$
collisions, the MPI and color reconnection (CR) seem to play an important role in deciding the particle production. PYTHIA8 with MPI and CR observed to explain various observables like $<p_T>$, charmonia production \cite{Thakur:2017kpv} etc. in  $p+p$ collisions at the LHC energies.
\section*{Strangeness Enhancement}
The enhancement of strangeness, which is quantified as the yield ratios of strange and multi-strange particles to the pion yields, is a signal of QGP. The high-multiplicity $p+p$ events seem to show a similar degree of strangeness enhancement like in A+A collisions.
 Figure~\ref{fig1} shows the $p_T$-integrated yield ratios of strange particles to pions $(\pi^++\pi^-)$ as a function of charged particle multiplicity
 density, $(dN_{\rm ch}/d\eta)$ at the mid-rapidity, $|y| <$ 0.5, as measured by the ALICE experiment at the LHC.  
 
\section*{The Nuclear Modification Factor}
\begin{figure}
\includegraphics[width=48mm]{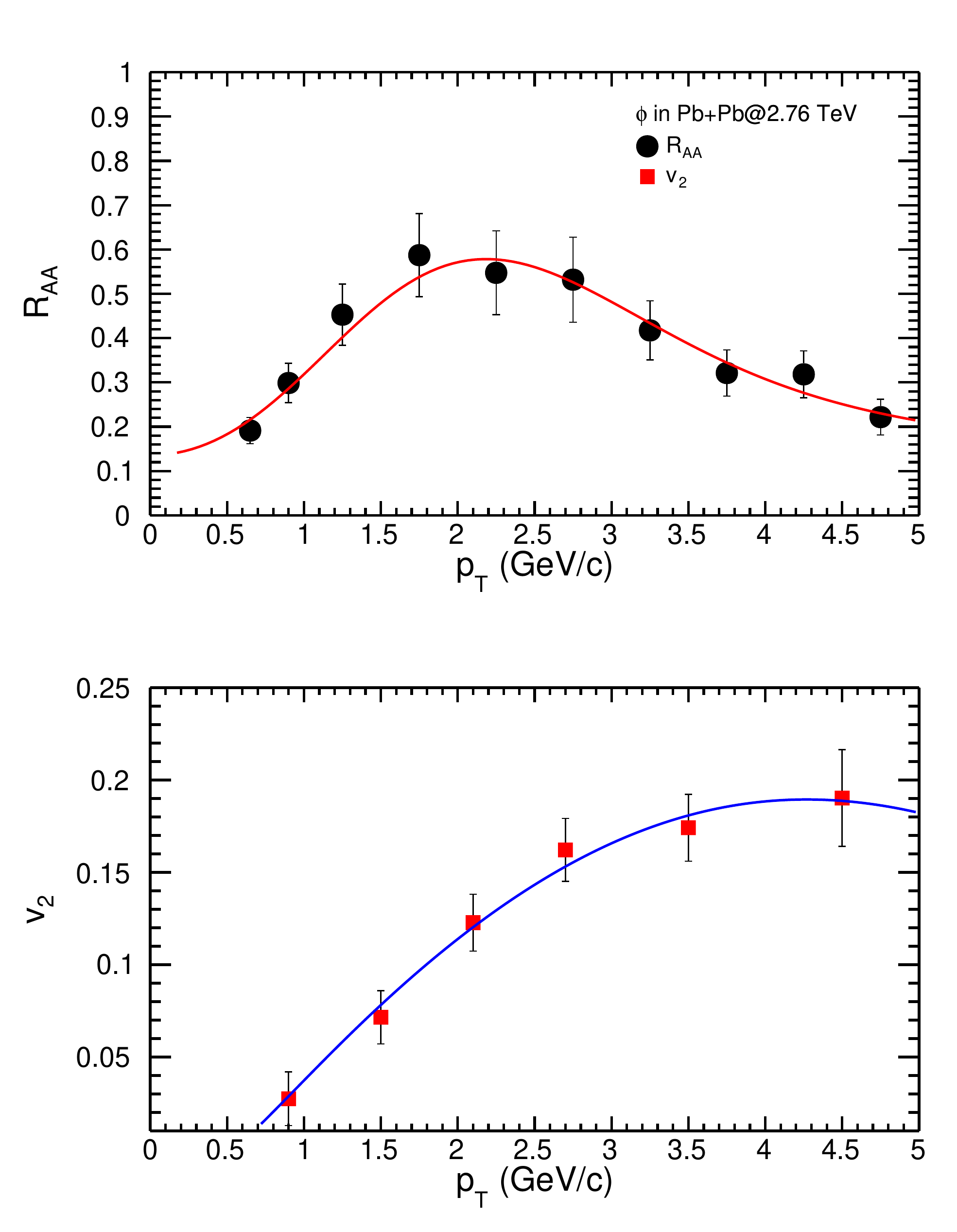}
\caption{\label{fig2}  The $R_{\rm AA}$ and elliptic flow, $v_2$ of $\phi$-mesons in Pb+Pb collisions at $\sqrt{s_{\rm NN}} = 2.76$ TeV \cite{Tripathy:2017kwb,Tripathy:2017nmo}.}
\end{figure}

Nuclear modification factor ($R_{\rm AA}$) is a measure of possible medium effects in A+A collisions. $R_{\rm AA} < 1$ suggests a suppression
of particle yields because of formation of a strongly interacting medium. A $p_T$ dependent study reveals different mechanisms like effect of
recombination (coalescence), jet-medium effects etc. Connecting the particle production in $p+p$ to heavy-ion collisions is one of the major challenges because of the different mechanisms involved. We have observed that the $R_{\rm AA}$ \cite{Tripathy:2017kwb} and elliptic flow \cite{Tripathy:2017nmo} in Pb+Pb collisions at the LHC are successfully explained by a formalism, which involves initial non-extensive $p_T$-distribution in $p+p$ collisions as an input to the Boltzmann Transport Equation (BTE) in Relaxation Time Approximation (RTA) with Boltzmann-Gibbs Blast Wave equilibrium distribution function.  Figure~\ref{fig2} shows the $R_{\rm AA}$ and elliptic flow, $v_2$ of $\phi$-mesons in Pb+Pb collisions at $\sqrt{s_{\rm NN}} = 2.76$ TeV. The  solid lines are the expectations from the discussed formalism, which seem to explain the experimental data \cite{Tripathy:2017kwb,Tripathy:2017nmo}.

\end{document}